# Spin wave mediated unidirectional Vortex Core Reversal by Two Orthogonal Monopolar Field Pulses: The Essential Role of Three-dimensional Magnetization Dynamics

Matthias Noske,[1,*] Hermann Stoll,[1] Manfred Fähnle,[1] Ajay Gangwar,[2] Georg Woltersdorf,[3] Andrei Slavin,[4] Markus Weigand,[1] Georg Dieterle,[1] Johannes Förster,[1] Christian H. Back[2] and Gisela Schütz[1]

[1]*Max Planck Institute for Intelligent Systems, Heisenbergstr. 3, 70569 Stuttgart, Germany*
[2]*University of Regensburg, Department of Physics, Universitätsstraße 31, 93053 Regensburg, Germany*
[3]*University of Halle, Department of Physics, Von-Danckelmann-Platz 3, 06120 Halle, Germany*
[4]*Oakland University, Department of Physics, Rochester, Michigan 48309, USA*

Scanning transmission x-ray microscopy is employed to investigate experimentally the reversal of the magnetic vortex core polarity in cylindrical $Ni_{81}Fe_{19}$ nanodisks triggered by two orthogonal monopolar magnetic field pulses with peak amplitude $B_0$, pulse length $\tau = 60$ ps and delay time $\Delta t$ in the range from $-400$ ps to $+400$ ps between the two pulses. The two pulses are oriented in-plane in the x- and y-direction. We have experimentally studied vortex core reversal as function of $B_0$ and $\Delta t$. The resulting phase diagram shows large regions of unidirectional vortex core switching where the switching threshold is modulated due to resonant amplification of azimuthal spin waves. The switching behavior changes dramatically depending on whether the first pulse is applied in the x- or the y-direction. This asymmetry can be reproduced by three-dimensional micromagnetic simulations but not by two-dimensional simulations. This behavior demonstrates that in contrast to previous experiments on vortex core reversal the three-dimensionality in the dynamics is essential here.

The zero field magnetic ground state of a flat cylindrical, soft ferromagnetic nanodisk (thickness $h$ of a few tens of nm and diameter $d = 2R$ of several hundred nm) is the vortex state. In the plane of the disk the magnetization curls in a clockwise (CW) or counter clockwise (CCW) circulation ($c = -1$ or $c = +1$). The exchange energy is minimized by the formation of a region with a typical diameter of 10 to 20 nm at the center of the disk where the magnetization turns out of the plane. This vortex core points up or down corresponding to the two polarities $p = +1$ and $p = -1$. The polarity can be considered as binary data bit, e.g., 1 for $p = +1$ and 0 for $p = -1$. Since the vortex polarity is very stable and well controllable, such structures may be used in the future for information storage and data processing devices. It is well established that the vortex polarity can be switched by applying dynamical external magnetic in-plane fields [1-6] or dynamical spin-polarized currents [7]. For excitation, short sinusoidal bursts [1] or unipolar pulses [2,8] can be used. Unidirectional switching which is a prerequisite for data storage applications is possible with rotating fields in resonance either with the gyrotropic eigenmode $G_0$ of the vortex core [3] with frequencies in the range of 100 MHz to 1 GHz (depending on the disk dimensions and the material of the disk), or with azimuthal spin waves in the higher GHz frequency range [4]. Alternatively, the excitation can consist of short multi-GHz rotating field bursts [5], or can have the shape of two orthogonal monopolar pulses [6] with pulse amplitude $B$, pulse length $\tau$ and a delay $\Delta t = \tau/2$ between the pulses. For the latter type of excitation, switching times shorter than 100 ps were found [6] for pulse amplitudes of a few tens of mT. The fact that digital pulses can be used for fast switching instead of rotating field bursts is an attractive aspect for potential technological applications.

---

[*]Corresponding author: noske@is.mpg.de

For all these types of excitations the dynamic switching mechanism is in general the same (see ref. [9] and references therein). Due to the excitation a local 'dip' region [10] is formed close to the vortex core where the out-of-plane components of the magnetization are antiparallel to the out-of-plane components of the core. If the excitation is strong enough, this dip splits into a vortex-antivortex pair with polarities opposite to the one of the original vortex, and the switching is performed via annihilation of the antivortex with the original vortex. This behavior has been shown by micromagnetic simulations [2]. In previous investigations on vortex core reversal the general results were very similar for two-dimensional [1,3-5] and for three-dimensional [2,6] simulations, showing that the three-dimensionality of the magnetization dynamics has an effect on the details of the reversal mechanism [9] but has only minor influence on whether the core is switched or not.

In the present paper we report on experimental results for the switching of the vortex core by two orthogonal monopolar in-plane magnetic field pulses in x- and y-direction with delay times $\Delta t$ between the two pulses which are considerably longer than the half width of the pulses. We explore experimentally the vortex core switching behavior to investigate unidirectional vortex core reversal. Therefore the phase diagrams for vortex core reversal as a function of pulse amplitude $B_0$ and $\Delta t$ are are studied. We compare two-dimensional and three-dimensional simulations with the experimental findings to demonstrate the importance of a three-dimensional treatment of the problem.

Cylindrical disks composed of Permalloy ($Ni_{81}Fe_{19}$) with height $h = 50$ nm and diameter $d = 500$ nm are thermally evaporated on top of cross-like copper (150 nm thick) strip lines with a width of 1.6 µm. Oxidation of the Permalloy disks is avoided by a 4 nm thick Al capping layer. The samples are prepared in multiple steps on top of silicon nitride membranes using a combination of optical and electron-beam lithography and lift-off processes. Magnetic field pulses of up to 35 mT are generated by sending current pulses through the crossed striplines (see Fig. 1 a). The excitation sequence consists of two Gaussian-like orthogonal magnetic pulses with a fixed pulse length $\tau = 60$ ps and a delay time $\Delta t = t_x - t_y$ between -400 ps and +400ps separating the pulses (see Fig. 1b and c). If the first pulse is in x-direction (y-direction), we attribute a negative (positive) value to $\Delta t$. The magnetic vortices are investigated by scanning transmission x-ray microscopy at the MAXYMUS endstation at BESSY II, Berlin, which provides a lateral resolution of 25 nm and a temporal resolution of 45 ps. The x-ray magnetic circular dichroism effect at the Ni $L_3$-edge is used as a contrast mechanism. Details of the measurement technique are given in ref. [6]. The switching phase diagram is determined by measuring the core polarity of the vortex structure before and after the pulse excitation.

The corresponding micromagnetic simulations are based on the Landau-Lifshitz-Gilbert equation [11], using the object-orientated micromagnetic framework [12] (OOMMF). Thereby the simulation is discrete in space and time. For three-dimensional simulations we use cubic simulation cells with a lattice constant of 3.125 nm well below the exchange length $L_{ex} = 6.6$ nm, for the two-dimensional simulations cells with a volume of $(3.125 \text{ nm})^2 \cdot h$ are used. Standard Permalloy material parameters are used for the exchange constant $A = 13 \cdot 10^{-12}$ J/m, gyromagnetic ratio $\gamma = 2.21 \cdot 10^5$ m/As, and anisotropy constants are set to zero. The damping parameter $\alpha = 0.007$ and the saturation magnetization $M_S = 690$ kA/m were determined as in ref. [6].

Fig. 2 shows the experimental phase diagrams for the vortex core switching (a: initial polarity 'down', $p = -1$, b: initial polarity 'up', $p = +1$) as function of the pulse amplitude $B$ and the delay time $\Delta t$. The phase diagrams are not symmetric for positive and negative delay times $\Delta t$. However, the switching behavior of both core polarities is nearly symmetric with respect to $\Delta t = 0$. This allows for unidirectional vortex core reversal in a wide range of pulse parameters, where switching occurs for one core polarity only. The symmetry between core 'up' and core 'down' is expected since the sense of rotation of each eigenmode of a magnetic vortex depends on the core polarity. Therefore the inversion of core polarity corresponds to an inversion of the delay time $\Delta t$.

Due to the symmetry between core 'up' and 'down', only the switching diagram for a vortex core 'up' (Fig. 2b) is discussed in the following. For negative $\Delta t$ and for positive $\Delta t$ up to 30 ps there are regimes where switching occurs, whereas for longer positive $\Delta t$ no switching is found. The switching threshold is minimal at $\Delta t \approx -310$ ps, $\Delta t \approx -150$ ps and $\Delta t \approx -30$ ps.

The phase diagrams obtained by two-dimensional and three-dimensional micromagnetic simulations are presented in Fig. 2c and 2d. Only the results for a vortex with initial polarity 'up' are shown as the simulation results for an initial core 'down' are perfectly symmetric as discussed above. The phase diagram from the two-dimensional simulations shows regimes with switching for positive delay times, where there is no switching in the experiment. Also at $\Delta t \approx -270$ ps there is a minimum in the switching threshold in the 2D simulations while the threshold is large in the experiment. In contrast to the two-dimensional case, in the three-dimensional simulations the location of the minima in switching threshold corresponds to the experimental results. However, also in the 3D simulations the switching regions are somewhat sharper compared to the experiment and have a higher switching threshold. Possible reasons for this discrepancy may be thermal excitations which are neglected in the simulations or sample imperfections which could act as nucleation centers for the switching process [13]. A further investigation of the time-resolved simulations reveals double-switching events (i.e. a succession of switching and back-switching) in the 2D case while no double switching events are found in the 3D case. Since the core polarity is the same as prior to the excitation, double switching events cannot be detected in the experiment.

In previous works [3,4,6,14] a sense of rotation of the external excitation field was defined. For pulsed excitation a CW (CCW) sense of rotation corresponded to positive (negative) delay times [6,14]. However, for delay times larger than half a period of the excited modes, it is no longer possible to attribute a specific sense of rotation to positive or negative delay times, as will be discussed in more detail below. We will discuss first the total energy pumped into the spin system by the excitation and second the effect of the vortex core trajectory. The total energy $E(t)$, which contains the demagnetization, exchange and Zeemann energy, has been calculated by 3D micromagnetic simulations. The maximum value max$[E(t)]$ that occurs for a given excitation with pulse amplitude $B_0$ and delay time $\Delta t$ is shown in Fig.3. The energy coupled into the system shows an oscillatory dependence on the delay time $\Delta t$ and switching only occurs in regions where the total energy exceeds $62 \cdot 10^{-18}$ J. The modulation of the injected energy can be explained by considering the lowest frequency spin-wave modes. These are the $n = 1, m = -1$ mode that rotates CW with an eigenfrequency of $f_{CW} = 6.45$ GHz and a period $T_{CW} = 155$ ps and the CCW rotating $n = 1, m = +1$ mode with $f_{CCW} = 8.62$ GHz and $T_{CCW} = 116$ ps. The eigenfrequencies were obtained by Fourier analysis of the magnetization dynamics after excitation with a short small-amplitude in-plane magnetic pulse. In the switching experiment, the first pulse excites both spin waves and the second pulse can either amplify or suppress these spin waves. The condition for amplification (compare sketches at the top of Fig. 3) is $\Delta t = \left(\frac{1}{4} + k\right) T_{CW}$ for the CW rotating mode and $\Delta t = \left(-\frac{1}{4} + k\right) T_{CCW}$ for the CCW rotating mode with an integer number $k$. The sketches also illustrate that the condition for amplification for both CW and CCW modes can be fulfilled for both, positive and negative delay times. The delay times where amplification occurs are marked by blue and green lines in Fig. 3, and for these delay time the largest values of max[$E(t)$] are found. This especially holds for the regions at $\Delta t \approx -310$ ps, $\Delta t \approx -150$ ps and $\Delta t \approx -30$ ps where minima in the switching threshold are observed (Fig.2). However, the simple consideration of energy injection is not sufficient to explain the experimental results. Constructive interferences and large energy values also occur for the positive delay times where switching is found only in 2D simulations (at $\Delta t \approx +80$ ps, $\Delta t \approx +200$ ps and $\Delta t \approx +330$ ps) but neither in the 3D simulations nor in the experiments.

We attribute the observed asymmetry in the phase diagram of the vortex core reversal to the 3D character of the eigenmodes which explains the failing of 2D simulations and a small additional excitation of the vortex gyrotropic mode. It is known, that the final stage of the vortex reversal, i.e. the annihilation of the antivortex with the original vortex, is in general a three-dimensional process [9]. As was discussed in the introduction, so far only minor differences between three- and two-dimensional simulations were found. In the following we show that here in contrast to previous investigations the situations in the 2D and 3D cases fundamentally differ even before the creation of the vortex-/antivortex pair, i.e. before the actual switching process begins. From the micromagnetic simulations we find that a strong z-dependent vortex core trajectory can be present before the start of the reversal process (see Fig. 4). This three-dimensionality changes the situation qualitatively and quantitatively and probably leads to differences in the vortex core reversal process: A moving vortex generates a gyrofield [15] that is responsible for the formation of the dip discussed in the introduction which may lead to a vortex core reversal. It is understandable that a z-dependent vortex core trajectory leads to a different gyrofield than a two-dimensional trajectory. This can be seen for example in Figs. 4c and 4d ($\Delta t = +330$ ps, $B_0 = 30$ mT), where a dip (blue region) is only formed in the two-dimensional simulations (4c) resulting in vortex core reversal ~30 ps after the presented snapshot, but only negligible dip-formation and no vortex core reversal occurs in the corresponding three-dimensional simulations (4d). The asymmetry between positive and negative delay times is most likely influenced by a small excitation of the vortex gyrotropic mode. Like for the spin-wave modes, a condition for optimal excitation of the CCW rotating gyrotropic mode exists according to $\Delta t =$ $\left(-\frac{1}{4}+k\right)T_{CCW,g}$ with the period $T_{CCW,g} = 1.7$ ns. For negative delay times, this condition can be first fulfilled for $\Delta t = -425$ ps, for positive delay times for $\Delta t = +1275$ ps. For this reason, in the experimental range of the delay time $-400$ ps $< \Delta t < +400$ ps, a more efficient excitation of the gyrotropic mode is expected for negative values of $\Delta t$. Indeed, the average gyration radius of the vortex core after the second pulse is slightly larger for negative delay times compared to positive delay times. This trend can be seen by comparing the trajectories in Fig. 4, where the average displacement of the vortex core in d) is approximately 20 nm larger compared to b) (switching occurs only for d). Obviously, only very little energy is coupled to the gyrotropic mode, as the energy values in Fig. 3 are comparable for positive and negative delay times. In spite of this general argument, the explicit relationship between the trajectories of the vortex core and the switching behavior is unclear. The concept of a critical velocity [15] cannot be applied here as velocities of approx. 600 m/s can be found for positive and negative delay times alike. We think that it is not possible to describe the switching behavior by a simple dynamical mechanism since the large amplitude excitation of multiple modes leads to non-linear coupling effects even before the switching process starts. Nevertheless the origin of the observed three-dimensional vortex core trajectories can be explained by the character of the vortex eigenmodes. To the knowledge of the authors, in the context of vortex core reversal so far the eigenmodes involved could always be treated in a two dimensional fashion, e.g., the gyrotropic mode and dipolar spin-wave modes. However, a two dimensional approach is no longer valid for the eigenmodes in vortex structures with film thicknesses above approximately 40 nm. In contrast, the dynamic profiles of the eigenmodes become three-dimensional as they can have nodes in the oscillation amplitude along the disk thickness. In particular the presence of higher-order gyrotropic flexure modes [16-19] $G_n$ leads to three-dimensional dynamics of the vortex core. Specifically, dynamic hybridization effects introduce a three-dimensionality (z-dependence) to the dynamics of the vortex core also for azimuthal spin-wave modes [20], which are mainly excited in the present experiments (see Fig. 3).

To conclude we have shown experimentally that unidirectional vortex core reversal is possible using two orthogonal monopolar magnetic field pulses of duration $\tau = 60$ ps over a wide range of amplitudes and delay times between the pulses. This flexibility might be useful in future data storage applications, for example to compensate for different signal propagation times. For this type of excitation, energy is mostly coupled to spin-wave modes, which modulates the observed switching threshold. Despite the fact that the vortex gyrotropic mode is excited only very little, it probably has a significant influence on the switching experiment. The experimental phase diagram for the reversal of the vortex core polarity can be reproduced only by three-dimensional micromagnetic simulations. This shows that the three-dimensionality of the magnetization dynamics is essential for this process and for this problem a two-dimensional simplification fails since the z-dependent vortex core dynamics are not captured.

The authors are indebted to Riccardo Hertel for valuable discussion. They gratefully acknowledge the technical support by Johannes Baumann, Michael Bechtel and Birgit Breimaier. The authors thank HZB for the allocation of synchrotron radiation beamtime.

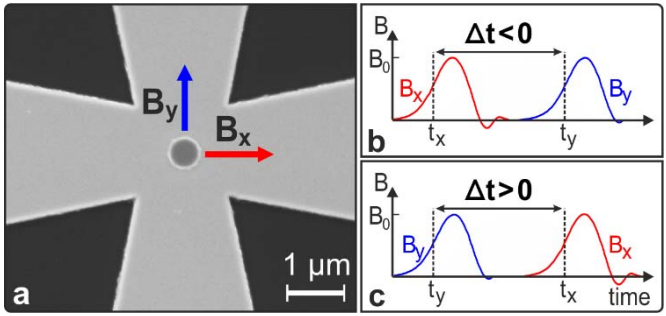

**FIG 1.** SEM image of a Permalloy vortex sample on top of crossed copper striplines (a) and excitation sequence for negative (b) and positive (c) delay times Δt consisting of two orthogonal magnetic pulses $B_x$ and $B_y$.

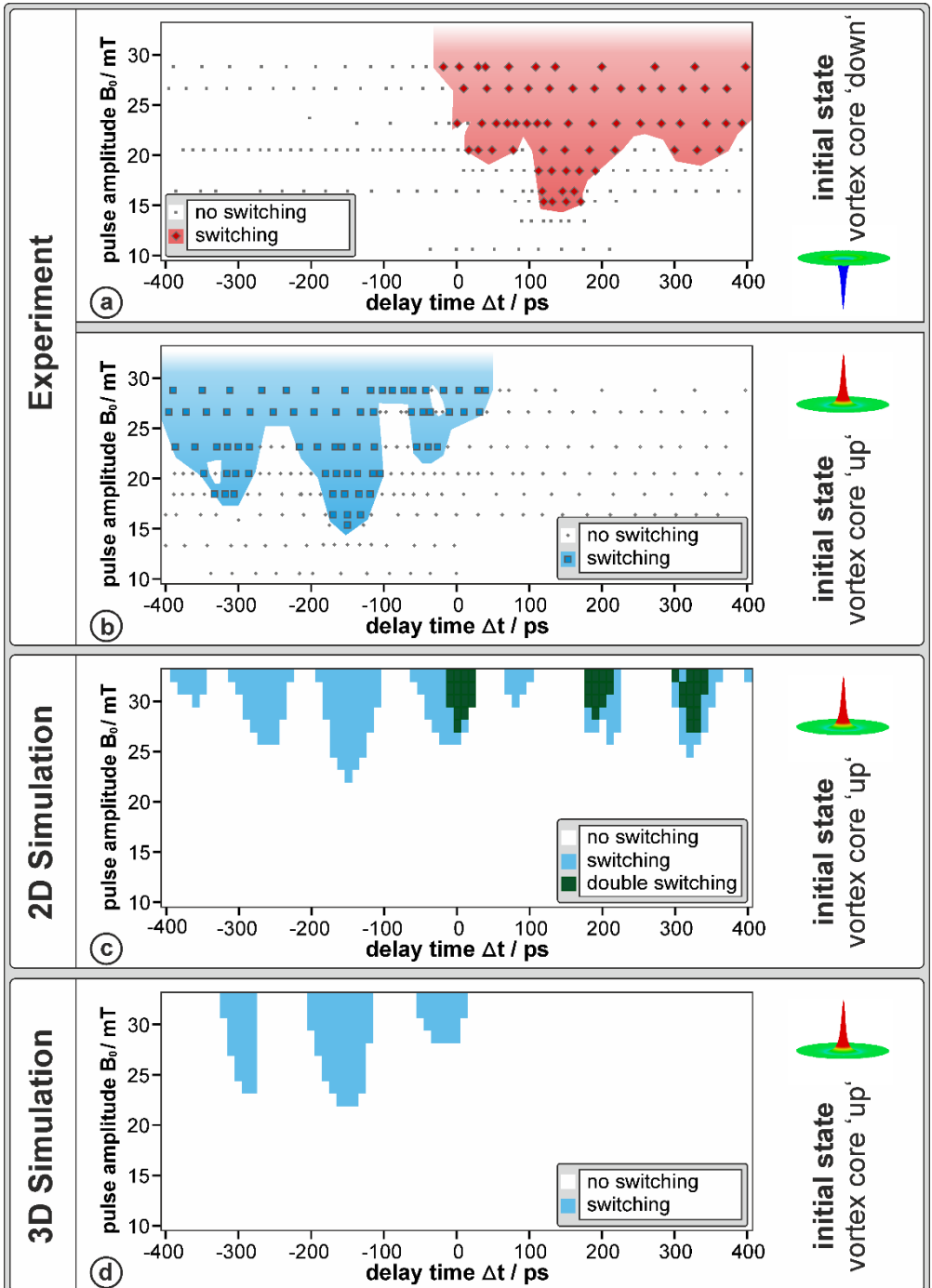

**FIG. 2.** Phase diagrams of vortex core reversal by two orthogonal in-plane magnetic field pulses of amplitude $B_0$, length $\tau = 60$ ps and delay time $\Delta t$ (see Fig. 1) between the pulses.

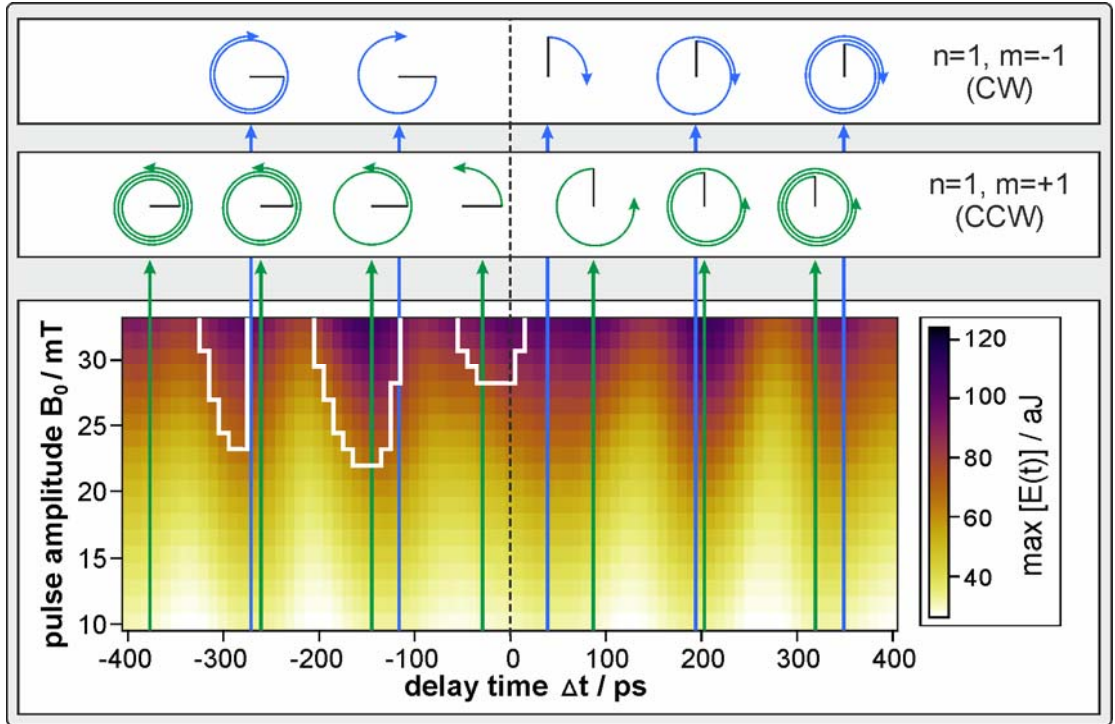

**FIG. 3.** The energy coupled into the vortex structure depends on the delay time between the two pulses, as can be seen by the maximum value of the total energy max[E(t)], which contains the demagnetization, exchange and Zeeman energy. The injected energy is largest for delay times where the CW and CCW spin waves are amplified by the second pulse as sketched at the top. Switching only occurs in regions of large energy: The white lines correspond to the switching threshold obtained from Fig. 2d.

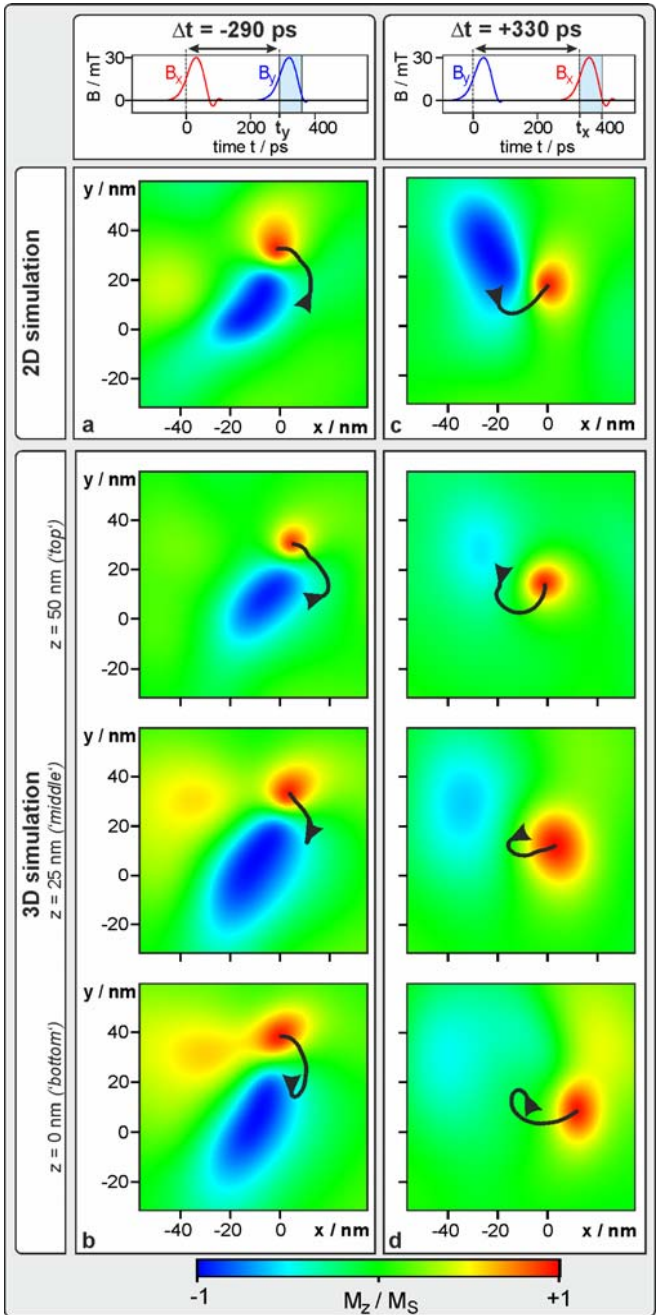

**FIG 4.** Snapshots of the z-component of the magnetization (vortex core up: red dot, 'dip' region: blue dot) and vortex core trajectories (black lines) to illustrate differences between 2D and 3D simulations and positive ($\Delta t = +330$ ps) and negative ($\Delta t = -290$ ps) delay times. The pulse amplitude is always $B_0 = 30$ mT. The trajectories are plotted from the start of the second pulse at $t_x$ (resp. $t_y$) until $t_x + 70$ ps (resp. $t_y + 70$ ps). The snapshots correspond to $t_x + 70$ ps (resp. $t_y + 70$ ps).

12